# Broadly-tunable smart glazing using an ultra-thin phase-change material

*Nathan Youngblood*[\*,1,§], *Clément Talagrand*[2], *Benjamin Porter*[1], *Carmelo Guido Galante*[3], *Steven Kneepkens*[3], *Syed Ghazi Sarwat*[1], *Dmitry Yarmolich*[4], *Ruy S. Bonilla*[1], *Peiman Hosseini*[2], *Robert Taylor*[5], *and Harish Bhaskaran*[\*,1]*

[1]Department of Materials, University of Oxford, UK
[2]Bodle Technologies Ltd., Begbroke, UK
[3]Eckersley O'Callaghan Ltd, London, UK
[4]Plasma App Ltd., Didcot, UK
[5]Department of Physics, University of Oxford, UK

[§]Current address: Department of Electrical and Computer Engineering, University of Pittsburgh, Pittsburgh PA, USA
[*]E-mail: *nathan.youngblood@pitt.edu* and *harish.bhaskaran@materials.ox.ac.uk*

**Abstract**

For many applications, a method for controlling the optical properties of a solid-state film over a broad wavelength range is highly desirable and could have significant commercial impact. One such application is smart glazing technology where it is necessary to harvest near-infrared solar radiation in the winter and reflect it in the summer—an impossibility for materials with fixed thermal and optical properties. Here, we experimentally demonstrate a smart window which uses a thin-film coating containing GeTe, a bi-stable, chalcogenide-based phase-change material which can modulate near-infrared absorption while maintaining neutral-colouration and constant transmission of light at visible wavelengths. We additionally demonstrate controlled down-conversion of absorbed near-infrared energy to far-infrared radiation which can be used to heat a building's interior and show that these thin-films also serve as low-emissivity coatings, reducing heat transfer between a building and its external environment throughout the year. Finally, we demonstrate fast, sub-millisecond switching using transparent electrical heaters integrated on glass substrates. These combined properties result in a smart window that is efficient, affordable, and aesthetically pleasing—three aspects which are crucial for successful adoption of green technology.

**Introduction**

Maintaining pleasant indoor temperatures of industrial and residential buildings consumes large amounts of energy, accounting for 20% to 40% of the national energy budgets in developed countries[1,2]. Carbon emissions associated with heating, cooling, and lighting both domestic and commercial buildings are particularly high in regions of the globe that experience large swings in the environmental temperature throughout the year—comprising over 12% of total $CO_2$ emissions in the UK[3,4], 14% in the US[5], and as much as 30% globally[6]. Windows account for a significant fraction of the external surface area of commercial buildings, and therefore energy loss; during colder months, a significant amount of heat is lost through windows in the winter season—as much as 25% in the US and UK and up to 50% in northern China[7,8]. In warmer months, on the other hand, windows transmit unwanted near-infrared solar energy which heat a building's interior, causing an unnecessary load on the cooling system. Upcoming legislation, such as the EU 2021 "nearly zero-energy buildings" regulation[9], will require smart solutions to this problem that provide significant energy gains without sacrificing aesthetic appeal.

Current static window solutions such as "Low-E" coatings can be used to reduce heat transfer[10], but cannot be actively switched to make use of near-infrared solar energy in winter months. Electrochromic smart windows have very slow response times (often several minutes)[11], require a continuous electric field in the "on" state[12], and cause attenuation over the entire solar spectrum upon switching[13,14]. Liquid Crystal (LCD) smart window technology switches much faster than electrochromics (several seconds), but modulates optical transmission via scattering and is therefore diffuse or tinted in the off state[15]. Photochromic and thermochromic smart windows, such as $VO_2$-based coatings, switch in a volatile manner[16], often have unwanted colouration[17], and can even incur net-energy costs to buildings in temperate climates[18]. Thus, the combination of manufacturing ease, bi-stability, and minimal changes in visible colouration has been elusive. Here, we experimentally demonstrate an active thin-film smart glazing that enables control over absorption of the near-infrared solar spectrum without an undesirable variation in opacity upon transition. Through photon-phonon energy down-conversion, we harvest near-infrared energy during the winter and re-emit it as far-infrared blackbody radiation. This down-conversion process enables additional benefits, such as anti-symmetric emission where absorbed energy is preferentially radiated into the building while preventing internal heat from escaping. Our unique design has significant implications for the future of smart windows and the design of energy efficient buildings.

**Results and Discussion**

The concept of our smart window design is illustrated in **Fig. 1a** where a smart glazing containing ZnS, Ag, and 12 nm of GeTe, a chalcogenide-based phase-change material (PCM), is applied to the outward-facing side of the internal window pane. This optical stack was optimized for maximum modulation of near-infrared solar energy while maintaining a constant transmission for visible light. The simulated optical spectrum for such a stack is shown in **Fig. 1c**. It is remarkable that such an active coating has a total thickness of less than 300 nm and can be optimized such that a significant fraction of the near-infrared solar energy is either reflected or absorbed depending on the state of the PCM. The external window pane in **Fig. 1a** is composed of uncoated glass and is added to protect the smart glazing and reduce the convective heat transfer between the outside environment and internal climate of the building. While this extra pane is recommended to improve the efficiency of the overall unit, it does not play a significant role in our window's optical response and we therefore limit our experimental efforts to the internal window containing the smart glazing.

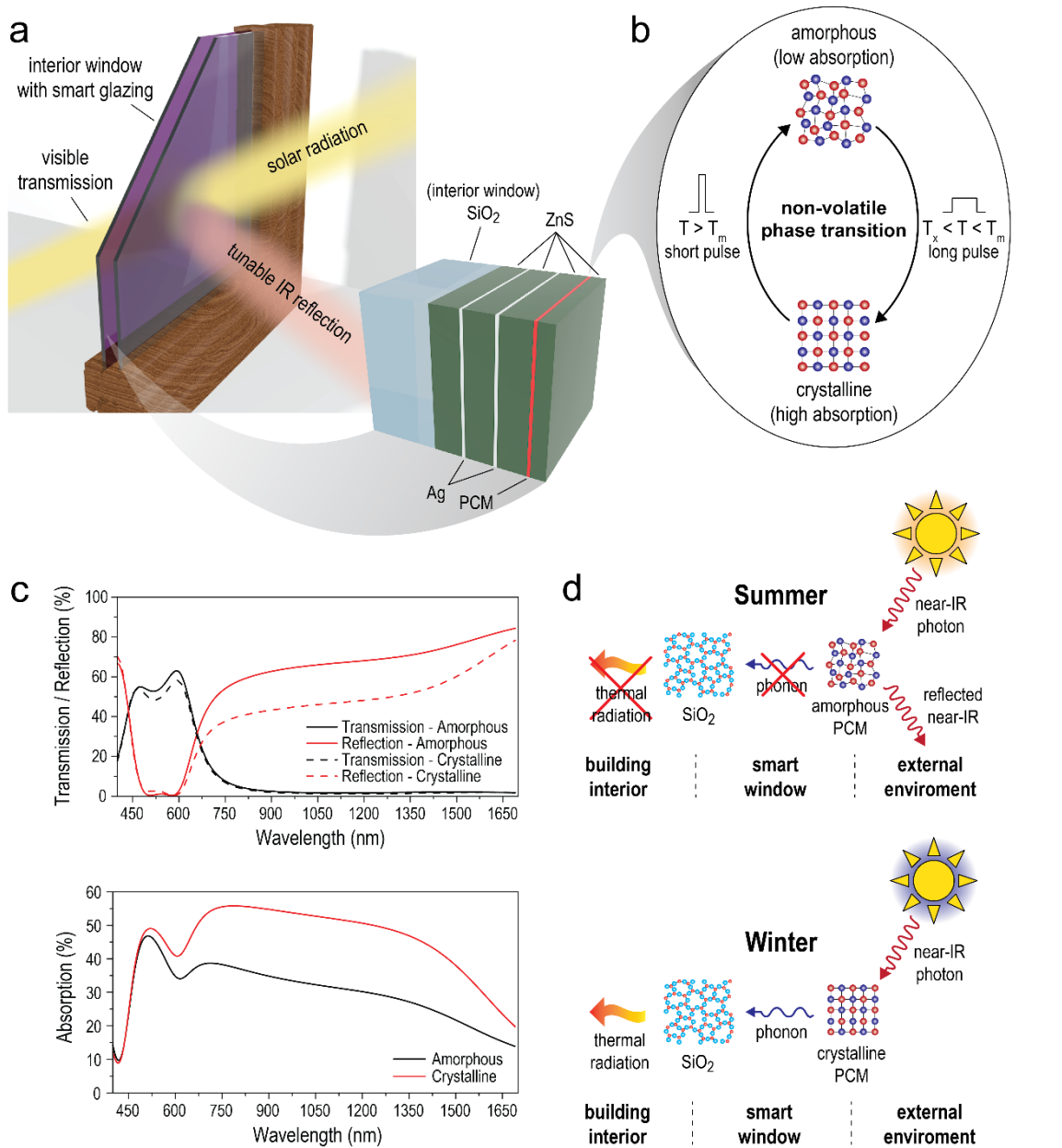

**Figure 1:** Phase-change smart glazing concept. **a)** 3D schematic of smart window design with interior window pane containing the smart glazing layer. Solar radiation is filtered and the near-infrared is selectively reflected or absorbed. Inset: Illustration of seven-layer smart glazing optical stack containing silver, ZnS, and a phase-change material (GeTe) layer. **b)** To initiate a phase-transition from crystalline to amorphous, a short thermal pulse greater than the melting temperature ($T_m$) is applied to the PCM layer via a resistive heater. A longer pulse between the crystallization ($T_x$) and melting temperature is used to return to the crystalline state. **c)** Simulated transmission, reflection and absorption of smart glazing. Optical spectra were optimized for minimizing change in visible spectrum while maximizing change of absorption at near-infrared wavelengths. **d)** Concept of thermal down-conversion used in our smart window design. In winter months, near-infrared radiation from the sun is absorbed in the smart glazing layer and converted to heat. In the summer months, the PCM is switched to the amorphous phase and reflects near-infrared photons back into the external environment.

To control the transfer of near-infrared solar energy through our smart window, we make use of the significant (and non-volatile) modulation of GeTe's complex refractive index when in the amorphous versus crystalline phase[19–23]. Here, both the real and imaginary components are modified upon crystallization resulting in increased absorption of the near-infrared spectrum. Thermally annealing the GeTe layer to a temperature above 450 °C for a short period of time before rapid quenching (less than 10 µs) randomizes the atoms in the lattice and results in an amorphous material with low near-infrared absorption (see **Fig. 1b**). Annealing for a longer time (greater than 10 µs) at a temperature of 280 °C allows GeTe to recrystallize, significantly increasing the absorption of near-infrared wavelengths. This process is non-volatile and reversible, meaning that energy is only consumed during the actual switching process and no electric field is needed to maintain either the amorphous or crystalline state[23]. Combined with the optimized Ag/ZnS optical stack, energy from the near-infrared is maximally absorbed or reflected while the transmitted visible light is minimally affected. In the crystalline state, the absorbed near-infrared energy is transferred to the $SiO_2$ glass window via phonons and re-emitted into the building through thermal radiation as illustrated in **Fig. 1d**. The high emissivity of $SiO_2$ and the relatively low emissivity of our optical stack (see **Fig. 3c-d**) allows for an efficient transfer of absorbed near-infrared energy to thermal radiation via this down-conversion process. Crucially, this process also allows our smart glazing to behave as a low-emissivity coating, which prevents far-infrared thermal radiation from entering or escaping the building when the PCM is in either the amorphous or crystalline state, which could potentially allow this active glazing to replace static low-e coatings while providing additional solar control.

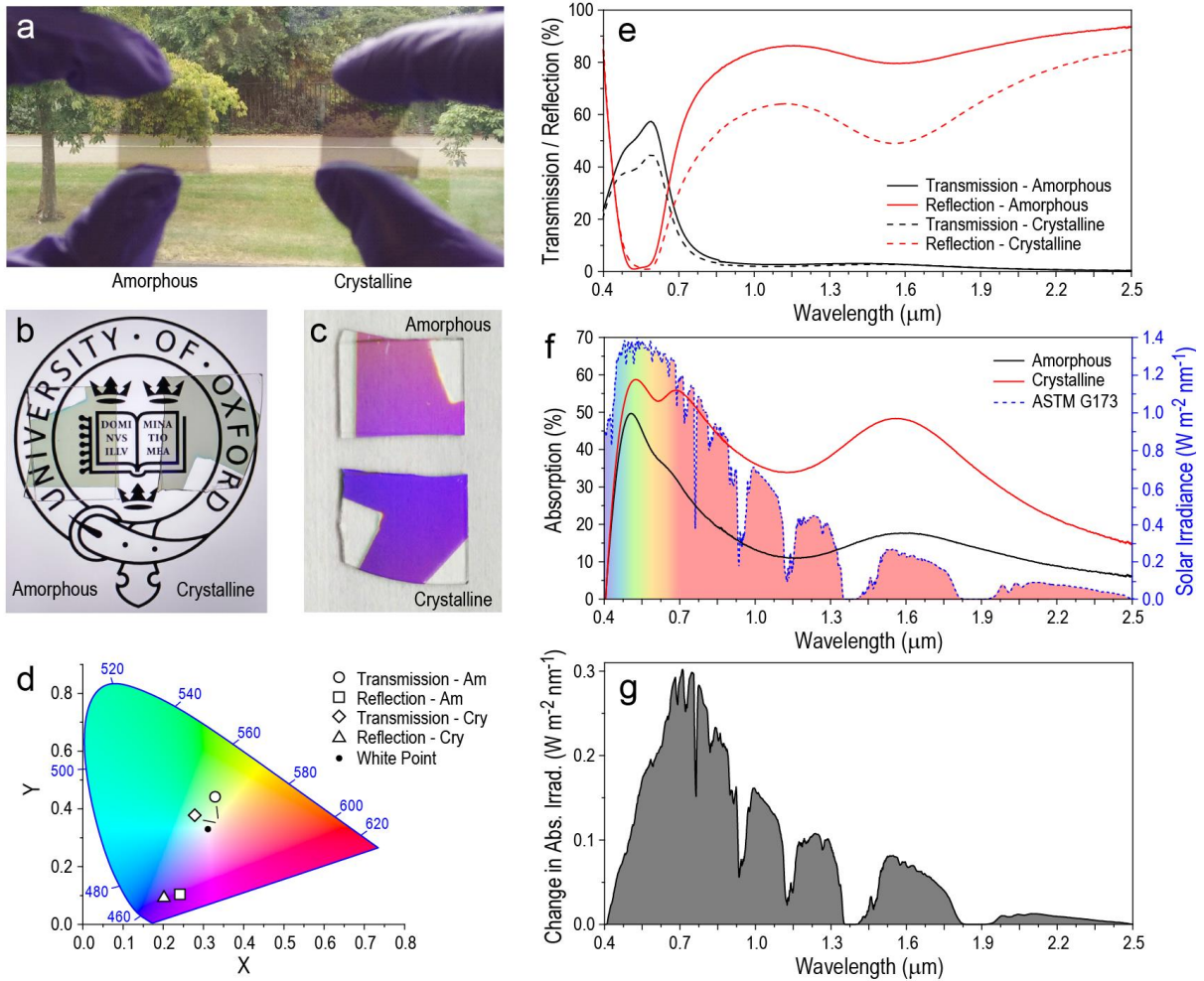

**Figure 2:** Experimental demonstration of phase-change smart glazing. **a)-c)** Optical images of transmission and reflection for fused quartz substrates with smart glazing on one side. **b)** Transmission of a white and black image from an LCD screen through samples with the smart glazing in both states. The optical transmission is very similar regardless of the state of the PCM layer. **d)** CIE colour map of experimentally measured transmission and reflection from samples shown in **a)**. **e)** Measured transmission and reflection spectrum of smart glazing in both the amorphous and crystalline state. **f)** Calculated absorption spectrum from **e)** plotted with the AM1.5 solar irradiance spectrum in blue. The change in absorption is broadband and covers the entire near-infrared region. **g)** Change in absorbed solar irradiance between the amorphous and crystalline states of the smart glazing.

Optical images of our fabricated smart windows are shown in both states in **Fig. 2a-c**. In **Fig. 2a** and **2b**. As expected, the transmitted visible light changes very little upon phase-transition. This is a highly desirable property over other technologies such as electrochromic windows where the significant change in visible transmission can cause unwanted disturbance to the building's occupants during a switching cycle[14]. Additionally, as quantified in the CIE 1931 chromaticity diagram shown in **Fig. 2d**, the colour of the transmitted light in both states remains very close to the white point, which represents a flat spectral power over all visible wavelengths. The reflected colour shown in **Fig. 2c** shifts slightly toward the blue end of the colour gamut upon crystallization as wavelengths at the red end of the spectrum are more strongly absorbed.

The optical response over the entire solar spectrum is visualized in **Fig. 2e** where we plot the transmission and reflection spectra of the smart glazing when the PCM is in both the amorphous and crystalline state. We observe relatively uniform transmission in the visible (approximately 50% transmission on average), while showing broadband modulation in the reflection of near-infrared wavelengths. Whilst the visible transmission of our smart glazing leaves room for improvement, we note that our values compare favourably with commercially available electrochromic smart windows in their fully transmissive state[13,14]. As the transmission of the near-infrared is below 5% in both states, we attribute the observed change in reflection to increased absorption when the PCM is in the crystalline state. We calculate the absorption spectra of our smart glazing in both states ($A = 1 - T - R$, where $A$, $T$, and $R$ are the absorption, transmission, and reflection spectra respectively) and plot them against the ASTM G173 solar irradiance standard[24] for an air mass of AM1.5g in **Fig. 2f**. Here we see that while the absorption is relatively high at visible wavelengths, the majority of absorption change occurs in the near-infrared. Crucially, the absorption in the near-infrared is low when the glazing is in the amorphous state which minimizes the amount of unwanted solar energy that is transferred to the building's interior during summer months. We plot the change in absorbed solar irradiance in **Fig. 2g** and show that the majority of the modulated energy occurs at wavelengths greater than 700 nm which have no benefit to illuminating a building's interior. Integrating the curves shown in **Fig. 2e-f** and weighting by the solar irradiance we find that 82% of unwanted near-infrared energy (780 nm to 2.5 µm) is reflected in the amorphous state while 42% is absorbed or transmitted in the crystalline state. This corresponds to a 130% increase in the amount of near-infrared solar energy allowed to enter a building by merely switching a 12 nm thick active phase-change layer.

To demonstrate that the absorbed near-infrared solar energy can indeed be down-converted to far-infrared thermal radiation, we designed an experiment to image the temperature increase of our smart window when the PCM is in the amorphous and crystalline states (see **Fig. 3a**). In this experiment, we used a FLIR® thermal camera to monitor the temperature of three samples: uncoated window (fused quartz), window with the smart glazing in the amorphous state, and window with the smart glazing in the crystalline state. The uncoated sample provided a reference thermal emission by which to calibrate the thermal camera and was subtracted from the final results shown in **Fig. 3b**. To illuminate the samples, we used an LOT, class ABA, solar simulator calibrated to 1 sun (1 kW/m$^2$) and an AM1.5g spectrum. A heat shield composed of plasterboard covered with thick aluminium foil was used to prevent unwanted thermal radiation from the simulator from saturating the thermal camera. An additional 1.3 µm thick sheet of polyethylene terephthalate (PET) was placed between the samples and the simulator which prevented additional convective heating or cooling of the samples while minimally influencing the transmitted solar spectra. **Fig. 3b** shows the temperature increase of the three samples measured every 30 seconds using the thermal camera. Both smart windows reached steady state shortly after 2 minutes of illumination with a relative 79 ± 5% increase in the temperature of the crystalline sample compared to the amorphous one. This agrees well with the percentage change (74%) in the total amount of solar energy (both visible and near-infrared) that is absorbed by the smart glazing according to the measured absorption spectra shown in **Fig. 2f**. We compared our experimental results to a simple thermal model in COMSOL Multiphysics® which included a uniform heat source and radiative and convective cooling boundary conditions. Using the convective heat transfer coefficient as a fitting parameter (12 W/m$^2$·K), we observe very good agreement between experiment and our simulation results (dashed lines in **Fig. 3b**).

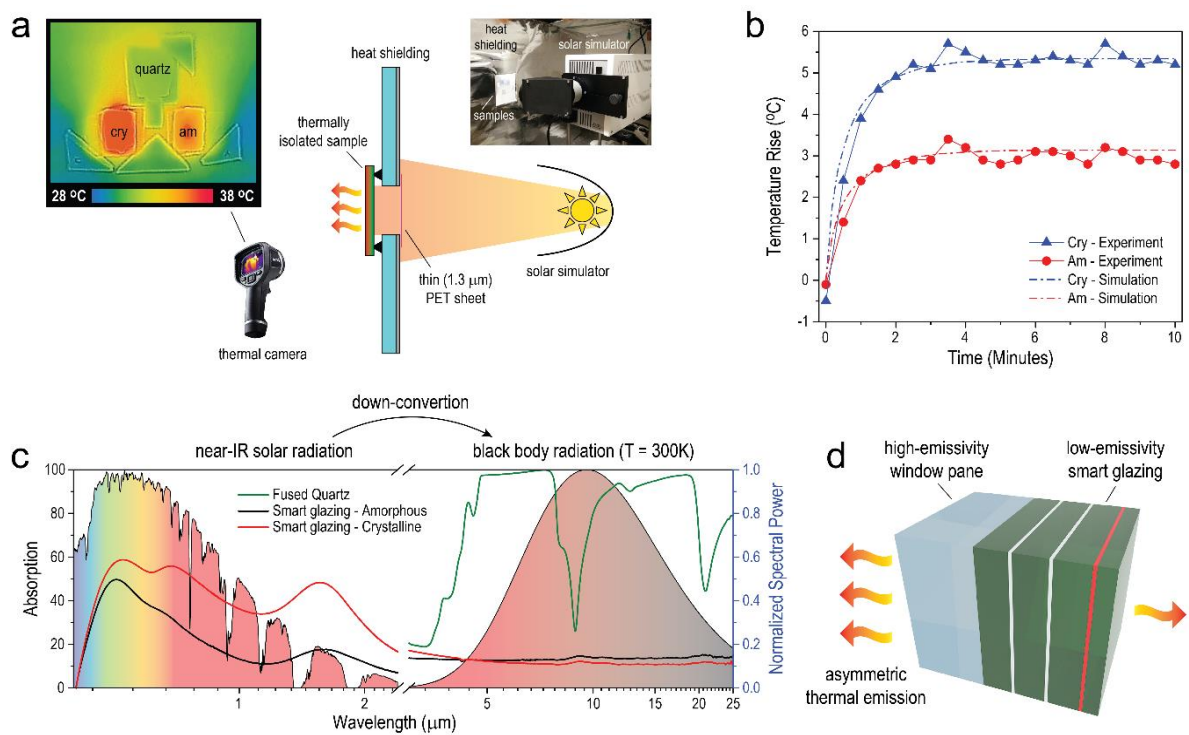

**Figure 3:** Demonstration of down-conversion of solar energy to thermal radiation. **a)** Experimental setup used to measure temperature increase in smart glazing through near-infrared absorption. A thermal camera is used to simultaneously measure the temperature of three samples (uncoated window and windows with smart glazing in the amorphous and crystalline state) as a function of time. **b)** Measured time-dependent temperature rise in the amorphous and crystalline samples. COMSOL® thermal simulations (dashed lines) agree well with experimental results. **c)** Spectral absorption of our smart windows ranging from the visible to far-infrared as measured from the glazed and unglazed surfaces. Measurements in the range of 3 µm to 25 µm were performed using FTIR and show a significant difference in the smart glazing versus uncoated side. **d)** Illustration of asymmetric thermal emission from window with smart glazing. The low-emissivity glazing prevents harvested near-infrared energy from re-emitting back into the external environment while reducing unwanted heat transfer through the windows regardless of the PCM state.

An additional benefit gained from down-converting the solar spectrum to far-infrared radiation is the asymmetric transfer of absorbed near-infrared energy through our smart window as illustrated in **Fig. 3d**. As our smart glazing is highly reflective to infrared radiation beyond 2.5 µm, we minimize the amount of harvested thermal energy which can be radiated back into the environment. On the other hand, SiO$_2$ performs as a highly emissive surface at temperatures around 300 K and efficiently re-emits absorbed near-infrared radiation into the building's interior. This asymmetry can be seen from the absorption spectrum in **Fig. 3c** measured on both the glass and smart glazing surfaces using Fourier-transform infrared spectroscopy (FTIR) for wavelengths between 3 µm and 25 µm. We overlay the absorption spectra with the theoretical emission spectrum of a black body at 300 K and observe significant overlap with far-infrared absorption from the glass side of the window. This overlap could be even further enhanced by adding an additional high-emissivity coating to the glass as demonstrated previously[25,26]. The absorption from the smart glazing side, however, remains below 20% over the entire far-infrared region in both the crystalline and amorphous states, resulting in a low-emissivity coating. This is a distinct advantage over other absorption-based smart window technologies (e.g. thermo- and electrochromics) which require the addition of a low-emissivity coating to prevent absorbed solar energy from thermally radiating into the building). Due to the asymmetric emissivity profile of our smart window, harvested near-infrared radiation will be down-converted through absorption and preferentially re-emit into the building when the PCM is in the crystalline state (**Fig. 3d**). Additionally, the low-emissivity smart glazing prevents thermal radiation from escaping or entering a building regardless of the state of the PCM, providing further energy savings throughout the year.

In a final experiment, we demonstrated electrically-controlled switching of the smart glazing by depositing our thin film stack on patterned FTO/glass substrates. For these devices, an improved PCM alloy was used which has similar optical properties to GeTe, but with a lower amorphization temperature[27]. **Fig. 4a** shows the experimental setup used to observe the switching behaviour of the smart glazing pixels in real time. A broadband light source (NKT WhiteLase Micro) filtered between 1.5 and 1.6 µm using dichroic and bandpass filters was used to generate optical power in the near-IR. The near-IR light was then coupled to an electro-optic probe station capable of sending electrical pulses while measuring the change in optical reflection. **Fig. 4b** shows optical microscope images of a typical device before and after initial electrical switching and after 100 switching cycles. Notably, the reflected visible light is consistent in colour and intensity throughout the switching events. The switching energy was measured using an in-series 25 Ω resistor and oscilloscope. The left side of **Fig. 4c** plots the extracted pulse waveforms and switching energy for the device shown in **Fig. 4b**. The right-hand plot in **Fig. 4c** shows the switching energy as a function of device area for multiple devices and pixel sizes. We see that while a lower peak power is required for crystallization, the total energy consumption is higher for crystallization rather than amorphization due to the longer pulse duration.

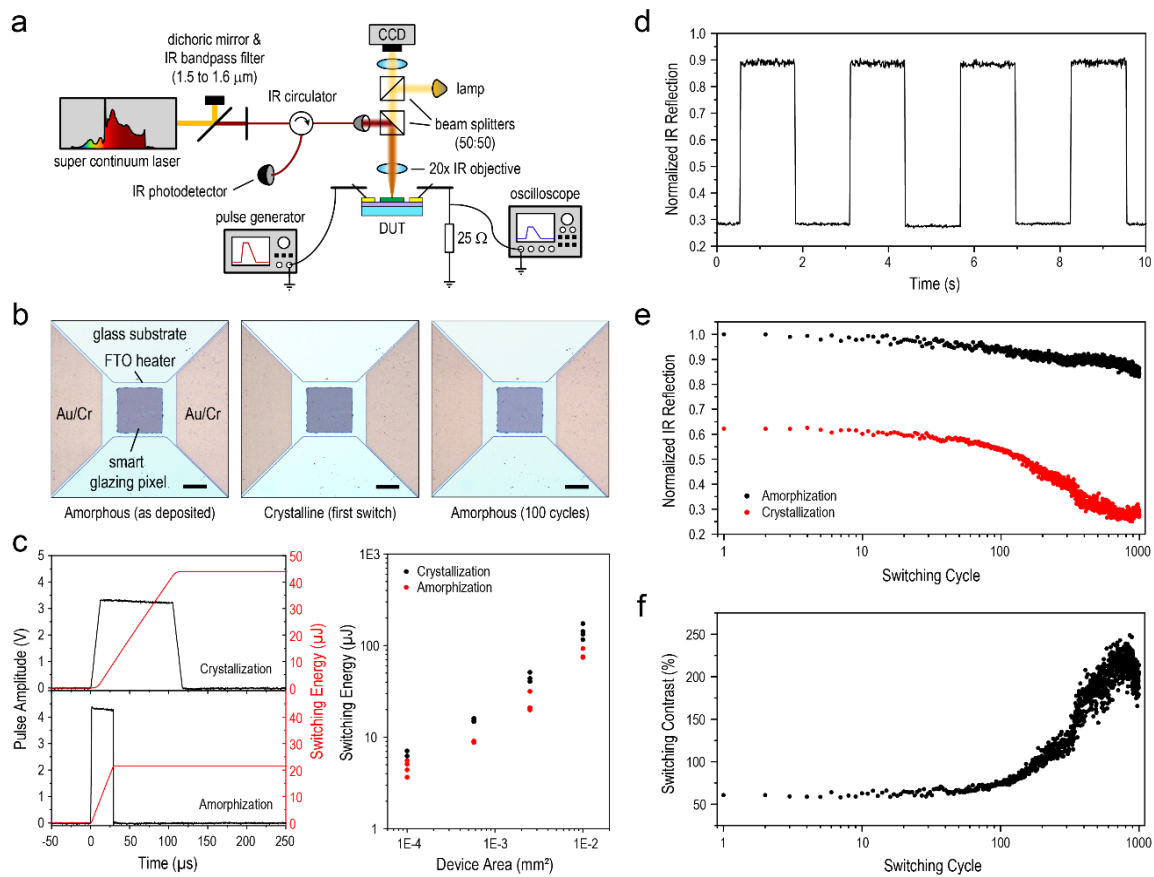

**Figure 4:** Electrical switching and cyclability of smart glazing pixels. **a)** Experimental setup used to switch the smart glazing state while measuring its reflection for wavelengths between 1.5 and 1.6 µm. **b)** Microscope images (using top and bottom illumination) of one device before switching, after first crystallization, and after 100 switching cycles (scale bar 25 µm). The visible reflection of the pixel does not vary noticeably during switching. **c)** Typical pulse shapes used to achieve crystallization and amorphization in the PCM thin film (left). Switching energy across multiple devices as a function of pixel area showing how energy scales with pixel size. **d)** Time trace of smart glazing pixel showing large near-IR contrast with fast (less than 1 millisecond) switching times. **e)** and **f)** Cyclability and switching contrast for device from **b)** and **c)** showing increased contrast as nucleation sites are formed.

We also investigated the cyclability of our smart glazing by switching between the two states while monitoring the IR reflection. **Fig. 4d** shows a time trace for multiple switching cycles of the device from **Fig. 4b** (time trace taken starting at 500 cycles), while **Fig. 4e**-**f** plot the device's cyclability and contrast for 1000 continuous cycles. Notably, we see both a reduction in the amorphous state reflection and an increase in contrast between the two bi-stable states. The first we attribute to silver migration in the optical stack during thermal cycling resulting in a reduction in IR reflectivity[28]. The increased switching contrast, on the other hand, is likely due to a greater fraction of crystalline domains forming due to the creation of nucleation sites. Indeed, for PCM memory, a "conditioning" treatment is often employed to stabilize the crystalline domain formation and improve the switching efficiency and cyclability of the memory cell[29].

In conclusion, we have demonstrated a thin-film smart glazing capable of controlling the transfer of near-infrared energy through a window using an ultra-thin phase-change active layer. This glazing provided a 130% modulation of near-infrared solar energy while maintaining a uniform transmission and colour at visible wavelengths. By down-converting near-infrared to thermal radiation, we were able to demonstrate efficient harvesting of solar energy during winter months while retaining a glazing with desirable low-emissivity properties regardless of the PCM state. We then demonstrate a possible route toward integration onto glass windows using transparent electrical heaters with an initial cyclability count of 1000 cycles. Our results provide a novel and aesthetically appealing alternative to current smart window technologies and could play a major role in reducing the carbon footprint of residential and commercial buildings in the future.

**Methods**

*Sample Fabrication:*

Smart glazing films were deposited via RF/DC sputtering on double-side polished quartz wafers using a Kurt J. Lesker PVD 75 system. The base pressure was $5\times10^{-7}$ Torr while the working pressure was 3 mTorr during deposition. RF power was 100 W and 50 W for ZnS and GeTe respectively, while the DC power was 50 W for Ag. All targets were 2 inches in diameter. To ensure accurate deposition thicknesses, the film thickness was monitored in-situ using an optical monitor which was compared with a model of the optical stack, providing precise thickness control and repeatability.

*Measurement Setup:*

Transmission and reflection measurements from 400 nm to 2.5 µm were performed using a PerkinElmer Lambda 1050 UV/VIS spectrophotometer under ambient conditions. Amorphous spectra were taken of the smart glazing as deposited, while crystalline spectra were measured after a 270° C anneal on a hot plate for 15 seconds. Mid- to far-infrared absorption was measured using a Varian Excalibur FTS 3500 FTIR spectrometer operating from 3 to 25 µm. For thermal down-conversion characterization, a FLIR© ONE Pro camera was used to measure the temperature rise in the uncoated quartz versus the quartz with smart glazing in the amorphous and crystalline states. The samples were illuminated with an LOT, class ABA, solar simulator calibrated to 1 sun (1 kW/m$^2$) and an AM1.5g spectrum. A heat shield composed of plasterboard covered with thick aluminum foil was used to prevent unwanted thermal radiation from the simulator from saturating the FLIR© thermal camera. The transmission of the 1.3 µm thick sheet of PET was also calibrated using the PerkinElmer Lambda spectrometer and factored into the calculated total solar absorption.

**Acknowledgements**

This research was supported by EPSRC via grants EP/J018694/1, EP/M015173/1, EP/M015130/1, and EP/M022196/1 in the UK.

**Contributions**

All authors contributed to this work. NY strategized and carried out all the experiments. NY and HB conceptualized and developed the concept and wrote the manuscript. GSS carried out some of the dynamic experiments in the work. CGG and SK carried out the modeling work on energy savings. BFP and DY aided fabrication of transparent heaters. CT and PH deposited the films on the dynamic heaters. RSB and RT helped NY with spectroscopic measurements on the films.

**Conflicts of Interest**

HB holds shares and serves on the Board of Directors at Bodle Technologies Limited. NY, HB, PH, CT and GSS have pending patent applications related to this technology. HB is employed at the University of Oxford which is incentivized to have papers published in top journals.

**References**

1. Pérez-Lombard, L., Ortiz, J. & Pout, C. A review on buildings energy consumption information. *Energy Build.* **40**, 394–398 (2008).

2. Kamalisarvestani, M., Saidur, R., Mekhilef, S. & Javadi, F. S. Performance, materials and coating technologies of thermochromic thin films on smart windows. *Renew. Sustain. Energy Rev.* **26**, 353–364 (2013).

3. Ricardo Energy & Environment, . *Sector, Gas, and Uncertainty Summary Factsheets*. (2017).

4. Delay, T., Farmer, S. & Jennings, T. Building the future today. *Carbon Trust* (2009).

5. Kwok, A. G. & Rajkovich, N. B. Addressing climate change in comfort standards. *Build. Environ.* **45**, 18–22 (2010).

6. OECD/IEA. *Energy Technology Perspectives 2015*. (2015).

7. Building fabric guide. *Carbon Trust* (2018).

8. Rezaei, S. D., Shannigrahi, S. & Ramakrishna, S. A review of conventional, advanced, and smart glazing technologies and materials for improving indoor environment. *Sol. Energy Mater. Sol. Cells* **159**, 26–51 (2017).

9. *Directive 2010/31/EU of the European Parliament and of the council on the energy performance of buildings*. (Official Journal of the European Union, 2010).

10. Hammarberg, E. & Roos, A. Antireflection treatment of low-emitting glazings for energy efficient windows with high visible transmittance. *Thin Solid Films* **442**, 222–226 (2003).

11. Granqvist, C. G. Electrochromics for smart windows: Oxide-based thin films and devices. *Thin Solid Films* **564**, 1–38 (2014).

12. Wang, Y., Runnerstom, E. L. & Milliron, D. J. Switchable Materials for Smart Windows. *Annu. Rev. Chem. Biomol. Eng.* **7**, annurev-chembioeng-080615-034647

(2016).

13. Lee, E. Application issues for large-area electrochromic windows in commercial buildings. *Sol. Energy Mater. Sol. Cells* **71**, 465–491 (2002).

14. Zinzi, M. Office worker preferences of electrochromic windows: a pilot study. *Build. Environ.* **41**, 1262–1273 (2006).

15. Kim, G. W. *et al.* Next generation smart window display using transparent organic display and light blocking screen. *Opt. Express* **26**, 8493 (2018).

16. Babulanam, S. M., Eriksson, T. S., Niklasson, G. A. & Granqvist, C. G. Thermochromic VO2 films for energy-efficient windows. *Sol. Energy Mater.* **16**, 347–363 (1987).

17. Smith, G., Gentle, A., Arnold, M. & Cortie, M. Nanophotonics-enabled smart windows, buildings and wearables. *Nanophotonics* **5**, 55–73 (2016).

18. Ye, H., Meng, X. & Xu, B. Theoretical discussions of perfect window, ideal near infrared solar spectrum regulating window and current thermochromic window. *Energy Build.* **49**, 164–172 (2012).

19. Yamada, N., Ohno, E., Nishiuchi, K., Akahira, N. & Takao, M. Rapid-phase transitions of GeTe-Sb2Te3 pseudobinary amorphous thin films for an optical disk memory. *J. Appl. Phys.* **69**, 2849–2856 (1991).

20. Wuttig, M. & Yamada, N. Phase-change materials for rewriteable data storage. *Nat. Mater.* **6**, 824–832 (2007).

21. Hosseini, P., Wright, C. D. & Bhaskaran, H. An optoelectronic framework enabled by low-dimensional phase-change films. *Nature* **511**, 206–211 (2014).

22. Park, J.-W. *et al.* Optical properties of (GeTe, Sb2Te3) pseudobinary thin films studied with spectroscopic ellipsometry. *Appl. Phys. Lett.* **93**, 021914 (2008).

23. Broughton, B. *et al.* 38-4: Solid-State Reflective Displays (SRD ®) Utilizing Ultrathin

Phase-Change Materials. *SID Symp. Dig. Tech. Pap.* **48**, 546–549 (2017).

24. Reference Solar Spectral Irradiance: Air Mass 1.5. *American Society for Testing and Materials (ASTM) Terrestrial Reference Spectra for Photovoltaic Performance Evaluation* Available at: http://rredc.nrel.gov/solar/spectra/am1.5/.

25. Raman, A. P., Anoma, M. A., Zhu, L., Rephaeli, E. & Fan, S. Passive radiative cooling below ambient air temperature under direct sunlight. *Nature* **515**, 540–544 (2014).

26. Zhai, Y. *et al.* Scalable-manufactured randomized glass-polymer hybrid metamaterial for daytime radiative cooling. *Science (80-. ).* **7899**, 1–9 (2017).

27. Castillo, S. G. *et al.* 57-4: Solid State Reflective Display (SRD ® ) with LTPS Diode Backplane. *SID Symp. Dig. Tech. Pap.* **50**, 807–810 (2019).

28. Leftheriotis, G., Yianoulis, P. & Patrikios, D. Deposition and optical properties of optimised ZnS/Ag/ZnS thin films for energy saving applications. *Thin Solid Films* **306**, 92–99 (1997).

29. Loke, D. K. *et al.* Ultrafast Nanoscale Phase-Change Memory Enabled By Single-Pulse Conditioning. *ACS Appl. Mater. Interfaces* **10**, 41855–41860 (2018).